\documentclass[11pt,twoside]{article}

\usepackage{asp2006}
\usepackage{epsf}
\usepackage{lscape}
\usepackage{graphicx}

\begin{document}
\title{Searching for the First Galaxies}
\author{Steven L. Finkelstein} 
\affil{George P. and Cynthia Woods Mitchell Institute for Fundamental Physics and Astronomy, Department of Physics and Astronomy, \\Texas A\&M University, 4242 TAMU, College Station, TX 77843}
\affil{stevenf@physics.tamu.edu}

\begin{abstract} 
As some of the first known objects to exist in the Universe, Lyman alpha emitting galaxies (LAEs) naturally draw a lot of interest.  First discovered over a decade ago, they have allowed us to probe the early Universe, as their strong emission line compensates for their faint continuum light.  While initially thought to be indicative of the first galaxies forming in the Universe, recent studies have shown them to be increasingly complex, as some fraction appear evolved, and many LAEs appear to be dusty, which one would not expect from primordial galaxies.  Presently, much interest resides in discovering not only the highest redshift galaxies to constrain theories of reionization, but also pushing closer to home, as previous ground-based studies have only found LAEs at $z > 3$ due to observational limitations.  In this review talk I will cover everything from the first theoretical predictions of LAEs, to their future prospects for study, including the HETDEX survey here in Texas.
\end{abstract}

\section{Introduction}
As new and more powerful instruments and telescopes are developed, we now probe deeper into the Universe than ever before.  While once redshifts of a few tenths were difficult to observe, we can now routinely probe to $z = 4$--$7$, corresponding to a time when the Universe was only $\sim 1$\,Gyr old.  The goal is simple in principle.  When we look at the galaxies around us, we see a well defined morphological sequence \citep{hubble26}, extending out to $z\geq 1$.  We would like to understand how the first galaxies formed, and through what evolutionary means they evolved into the galaxies we see today.

To study the early Universe, we need techniques to accurately determine the distances to galaxies.  Locally, various distance ladder rungs are used, but the most powerful of these, SNe Ia, taps out at $z\sim 1$.  At high redshift, a common method is to use the expected spectral features of the galaxies we hope to find, such as with the Lyman break technique \citep{steidel93}.  At $z \geq 2$, the Lyman limit, an intrinsic spectral break at 912\,\AA, shifts into the observed optical.  Though the intergalactic medium (IGM) is largely ionized, enough remnant neutral gas is in a typical line-of-sight to efficiently absorb photons bluer than the Lyman limit, increasing the amplitude of this break.  Additionally, from $912$--$1216$\,\AA, a forest of absorption lines appear due to absorption of intervening H\,{\sc i} along the line of sight, known as the Ly$\alpha$~forest.  At higher redshifts, the number of absorbers increases, and the opacity in the Ly$\alpha$~forest increases dramatically, such that the spectral break begins at 1216\,\AA, and is called the Ly$\alpha$~break.

Using this spectral feature, one can discover large samples of Lyman break galaxies (LBGs) by observing with three broadband filters.  For example, at $z \sim 3$, the Lyman break is in the $u^{\prime}$-band, thus one would observe with $u^{\prime}$, $g^{\prime}$ and $r^{\prime}$ filters.  As the break passes through the $u^{\prime}$-band, the $u^{\prime}-g^{\prime}$ color will increase.  By examining how the colors of stellar population model tracks evolve with redshift, one can select a $u^{\prime}-g^{\prime}$ color which will select galaxies at $2.5 \leq z \leq 3.5$.  However, this single color can also include red galaxies at lower redshifts, thus one typically also demands a blue rest-frame ultraviolet (UV) color, $g^{\prime}-r^{\prime}$ in our example.  With this method, one can efficiently select large numbers of high-redshift star-forming galaxies\footnote[1]{See a visualization of this selection at: http://shepherd.physics.tamu.edu/select$\_$movie.html}.

However, the LBG selection method relies on robust continuum detections in multiple images, as well as a very deep dropout filter.  Thus, typical LBG studies place a brightness limit on their sample, typically $R\leq 25.5$ \citep{steidel96}, leaving a large population of faint galaxies unexplored.  Additionally, though galaxies selected with this method should lie in a redshift slice of $\Delta$$z \sim 1$, one requires followup spectroscopy to confirm the selection criteria and determine the redshift selection function.  This can be costly, as even ``bright'' LBGs ($m_{\rm AB}\sim 23$) can require many hours on 8--10m class telescopes to confirm their redshifts via absorption lines.

If the goal is to study the earliest galaxies, then one can rely on some additional spectral features.  In particular, they should be heavily star-forming, and thus bright in numerous nebular emission lines.  \citet{partridge67} predicted that one could select the first galaxies on the basis of their Ly$\alpha$~emission, which at a rest wavelength of 1216\,\AA~shifts into the optical from $2 \leq z \leq 6$.  With a typical initial mass function (IMF), many stars will form with sufficient mass (and thus temperature) to emit large quantities of hydrogen-ionizing photons ($\lambda \leq 912$\,\AA).  These photons will ionize H\,{\sc i} out to approximately the Str$\mathrm{\ddot{o}}$mgren radius.  At this point, there is a balance between ionization and recombination, and when a proton and electron recombine, a Ly$\alpha$~photon will be emitted $\sim 67$\% of the time \citep[Case B, ][]{osterbrock89}.  \citet{partridge67} predicted that $> 5$\% of the total luminosity of these galaxies could emit in this single line.

Spurred by these predictions, a number of studies searched for these Lyman alpha emitting galaxies (LAEs), but to no avail \citep[e.g.,][]{koo80,meier81}.  A number of causes were suggested to explain the lack of LAEs, such as the possibility that these galaxies could be dusty \citep[e.g.,][]{meier81}.  In a uniform interstellar medium (ISM), Ly$\alpha$~photons will resonantly scatter off H\,{\sc i}, resulting in longer path lengths, and thus they stand a much greater chance of being absorbed by dust.  Or, some reasoned that perhaps these objects simply didn't exist, or were far too faint to be observed.

Thankfully, fortunes turned in the late 1990's, as two groups searching for LAEs, \citet{cowie98} and \citet{rhoads00}, discovered large samples.  Why were they able to do this while previous groups were not?  The answer is a matter of technological advancement.  \citet{cowie98} used the new Keck 10m telescope, while \citet{rhoads00} used the new MOSAIC large format detector on the Kitt Peak National Observatory 4m telescope.  The deeper imaging afforded by Keck or the wider field-of-view provided by MOSAIC both resulted in a larger volume being probed, and thus gave these studies a greater chance of discovery.

These, and subsequent studies, used the method of narrowband selection to identify these galaxies on the basis of their bright Ly$\alpha$~lines.  To do this, one observes with a narrowband filter, where the redshift of the LAEs one finds is defined by the central wavelength and width of the filter.  For example, \citet{rhoads00} used a set of filters around 6500\,\AA, thus the LAEs they found were at $z \sim 4.5$.  One also observes the field with a broadband filter encompassing the narrowband filter.  Examining sources in both images, one selects objects with a flux excess in the narrowband image as possible emission line candidates.  However, low-redshift interlopers are possible; for example, with a narrowband filter at 6500\,\AA, one could also select [O\,{\sc ii}], [O\,{\sc iii}] or H$\beta$ emitters.  To screen against these, one typically enforces a dropout criterion similar to the Lyman break method, requiring an additional broadband image blueward of the line.

\begin{figure}[!ht]
\centering
\includegraphics[width=5in]{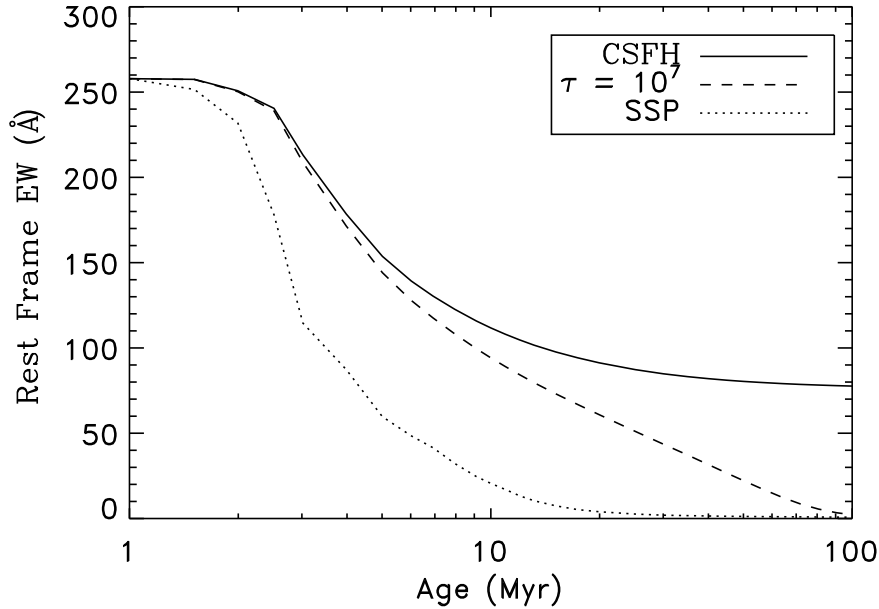}
\caption{Model variation of Ly$\alpha$~EW with age for three different star formation histories: constant, exponentially decaying, and a simple stellar population, or burst (from Finkelstein et al.\ 2008).  All of these models have Z = 0.02 Z$_{\odot}$, and include the effect of IGM absorption at $z \sim 4.5$, where we assume that half of the line flux (i.e.~the blue wing) will be subject to IGM attenuation.}\label{ewdist}
\end{figure}

Using this method of narrowband selection, \citet{cowie98} and \citet{rhoads00} both found that one could discover large samples of LAEs fairly easily, at least at redshifts where Ly$\alpha$~is observed in the optical.  One of the first interesting discoveries about LAEs was that their Ly$\alpha$~equivalent widths (EWs) were strong (the EW is a measure of the line strength relative to the continuum).  In many cases, the EWs exceeded predictions from normal stellar population models \citep[e.g.,][]{charlot93}.  Figure~\ref{ewdist} shows the distribution of Ly$\alpha$~EW versus age for different star-formation histories (SFHs).  Unless one is observing a LAE in the first few Myr, the Ly$\alpha$~EW should be $<$ 200\,\AA.  However, this is frequently not the case \citep[e.g.,][]{kudritzki00,malhotra02}.  Thus, something unexpected is occurring in many LAEs.

\section{Uncovering the Physical Properties of LAEs}
There are a number of physical possibilities which could result in these high EWs.  If LAEs are truly the first galaxies, or their immediate descendants, we might expect them to have very low, or perhaps zero metallicities.  The lack of metal lines for cooling during star formation can thus result in an initial mass function (IMF) which preferentially forms more massive stars than \citet{salpeter55}, or a so-called top-heavy IMF.  Heavier and more metal-poor stars have hotter stellar photospheres, which result in the production of more ionizing photons, and thus more Ly$\alpha$~photons.  To examine the possibility of this scenario will require detailed stellar population modeling or direct measurements of the metallicity from rest-frame optical emission lines.

If these galaxies host super-massive black holes which are actively accreting material, then much of the light we see could originate from the accretion disk around the black hole, including any Ly$\alpha$~emission.  The Ly$\alpha$~EW would thus not be subject to such strict limits as in the case of star formation.  If many LAEs hosted active galactic nuclei (AGN), this could solve the EW problem.  However, at high redshift the only means we currently have to diagnose the presence of AGN is via their X-ray emission.  Typically LAE studies examine X-ray data, and they find that the AGN contamination is low, from 1--5\% at 2 $<$ $z <$ 5 \citep[e.g.,][]{malhotra03, wang04, gawiser06b, nilsson09}.  Thus, AGN are likely not a major contaminant, although low-luminosity AGN may be lurking beneath current X-ray sensitivities \citep[see, e.g.,][]{finkelstein09e}.

A different scenario for the high EWs observed in LAEs involves a dusty ISM.  While dust was originally thought of as the reason why LAEs were \textit{not} seen in the 1980's, this relied on the assumption of a homogeneous ISM.  If, however, the ISM was clumpy, where neutral gas and dust were mixed together in clouds with a relatively empty inter-cloud medium, it is possible to actually enhance the Ly$\alpha$~we would measure \citep{neufeld91, hansen06}.  This can happen because Ly$\alpha$~photons are resonantly scattered by H\,{\sc i}, while continuum (and other emission lines) are not.  Continuum photons traveling through such an ISM would penetrate the surface of these clouds.  Some of these photons will travel straight through and escape the galaxy, some will be absorbed by dust, and some will be scattered by the dust into a different direction.  However, Ly$\alpha$~photons will resonantly scatter with the H\,{\sc i} on the surface of these clumps, essentially bouncing off.  Thus, Ly$\alpha$~photons stand a significant chance of evading interaction with dust in such an ISM.  This scenario has the net effect of increasing the ratio of the Ly$\alpha$-to-continuum escape fractions, and thus increasing the observed EW.

\subsection{Spectral Energy Distribution Fitting}
The typical method of uncovering the physical properties of distant galaxies is through spectral energy distribution (SED) fitting, whereby one compares observations of a galaxy (in most cases broadband photometry) to a suite of stellar population models \citep[e.g.,][]{papovich01, shapley01}.  One varies the age, dust extinction, metallicity, SFH, etc.\ of the models, and finds the one that best represents the observations, typically by minimizing the $\chi^{2}$ between the models and the observations.

The first SED analysis of LAEs hoped to answer the question of whether or not they truly were the primitive first galaxies.  However, many of these studies relied on ground-based data, thus they had to stack the fluxes of their sample together to obtain the necessary signal-to-noise (S/N) to robustly compare them to models.  \citet{gawiser06b} studied 18 spectroscopically confirmed LAEs at $z = 3.1$, finding that on average they had ages of 90\,Myr, masses of $5\times 10^{8} M_{\odot}$, and zero dust extinction.  In \citet{finkelstein07}, we stacked nearly 100 LAEs at $z \sim 4.5$ into three bins by EW strength, and found age and mass ranges from 1--40\,Myr, and 10$^{7-9}$ $M_{\odot}$, respectively, with the highest EW sources having the youngest ages and lowest masses.

In the first studies to individually analyze LAEs, \citet{lai07} and \citet{pirzkal07} found strikingly different results, although the differences can be explained by their selection biases.  \citet{lai07} studied LAEs at $z \sim 5.7$ in the Great Observatories Origins Deep Survey (GOODS) North field, restricting their analysis to three LAEs which were detected in the infrared (IR) with the \textit{Spitzer Space Telescope}.  They found that these objects had ages from 5--700\,Myr, masses from 5 $\times$ $10^{9-10} M_{\odot}$, and significant dust extinction from $A_V = 0.4$--$1.7$\,mag.  However, the restriction to \textit{Spitzer} detections likely biases towards more massive LAEs, though the existence of LAEs this massive is intriguing.  In a complementary study, \citet{pirzkal07} studied 10 LAEs individually at $4 < z < 5.5$ discovered in the {\it Hubble} Ultra Deep Field (HUDF) via grism spectroscopy with the Advanced Camera for Surveys (ACS) on board the \textit{Hubble Space Telescope} (\textit{HST}).  This method of discovery in the extremely small HUDF will likely bias this study towards intrinsically fainter and lower mass objects.  Correspondingly, \citet{pirzkal07} find ages of 1--20\,Myr, $M = 3\times 10^{7}$--$2\times 10^{9} M_{\odot}$ and extinction of $A_{V} = 0$--$0.6$\,mag.  Thus, the faintest LAEs are among the lowest mass and youngest galaxies in the Universe.

\subsection{Investigating the ISM Geometry}
However, these early studies did not use an important piece of information we have on these objects, namely that they exhibit Ly$\alpha$~emission.  Ly$\alpha$~can have a significant effect on the observed SED, increasing the broadband magnitude by up to 0.5\,mag for strong emitters \citep[e.g.,][]{papovich01}, thus Ly$\alpha$~emission should be included in the models.  Additionally, since the majority of LAEs were discovered via narrowband photometry, the narrowband data should be included in the fit.  However, most studies use the popular \citet[][hereafter BC03]{bruzual03} stellar population synthesis models, which do not include nebular emission lines.

\begin{figure}[!ht]
\centering
\includegraphics[width=5in]{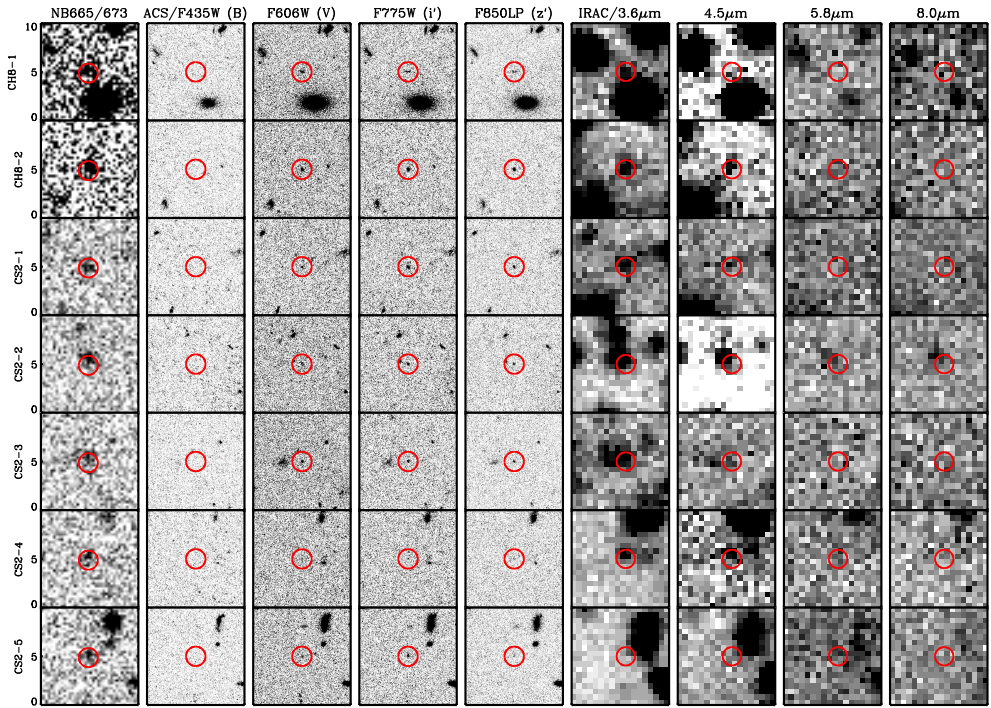}
\caption{10{\hbox{$^{\prime\prime}$}}~cutouts of seven of the $z \sim 4.5$ LAEs from (from Finkelstein et al.\ 2009a).  They are extremely bright in the narrowband and undetected in the B-band.  11/14 LAEs were detected in at least one {\it Spitzer} band.}\label{laestamps}
\end{figure}

This was the motivation behind our work in \citet{finkelstein08} and \citet{finkelstein09a}.  We obtained narrowband imaging from the Cerro Tololo InterAmerican Observatory 4m telescope, using the MOSAIC II wide-field imager in three filters: H$\alpha$ ($\lambda_{c}$ = 6563\,\AA), H$\alpha$+80 ($\lambda_{c}$ = 6650\,\AA) and [S\,{\sc ii}] ($\lambda_{c}$ = 6725\,\AA).  These filters can thus select LAEs at 4.37 $\leq$ $z \leq 4.57$.  We chose to observe the Extended Chandra Deep Field South (ECDF-S), which contains the GOODS-South field, consisting of extremely deep \textit{HST} and \textit{Spitzer} data.  We found 14 LAEs in the GOODS region, restricting our analysis only to objects in the deep {\it HST} fields so that we could analyze these LAEs individually.  Figure~\ref{laestamps} shows image cutouts of seven of our LAEs from this study.

To take advantage of the measured Ly$\alpha$~properties of these objects (from their narrowband excesses), we added Ly$\alpha$~emission to the models, basing the line strength off the number of ionizing photons at a given age, metallicity and SFH.  Adding in this emission also allowed us to examine the dust enhancement scenario \citep[e.g.,][]{neufeld91} discussed earlier, as we were able to attenuate Ly$\alpha$~by a different amount than the continuum when dust extinction was added to the models.  Our treatment was slightly more complex than simple clumpy spheres, as we allowed the amount that dust affected Ly$\alpha$~for a given model to be a free parameter, which we called ``$q$'', where $q$ is defined by:
\begin{equation}
f^{\prime}_{\rm Ly\alpha} = f_{\rm Ly\alpha} * \mathrm{exp}(-q\tau_{c})
\end{equation}
where $\tau_{c}$ is the optical depth of dust to the continuum around Ly$\alpha$.  A value of $q = 0$ thus symbolizes the Neufeld scenario, where Ly$\alpha$~escapes any dust attenuation, and $q = 10$ symbolizes a homogeneous ISM, where resonant scattering ensures that Ly$\alpha$~will be attenuated much more than the continuum.

When all of this was included in our models, we found the interesting result that LAEs are not a homogeneous population of galaxies.  Out of our sample of 14 galaxies, we found that 12 appeared to be young, with best-fit ages of 3--15\,Myr and masses of $0.16$--$6 \times 10^{9} M_{\odot}$.  However, the remaining two LAEs were much older, both with ages of $\sim 500$\,Myr, and masses of 0.4--$5 \times 10^{10} M_{\odot}$.  Perhaps more surprising is that all of the galaxies in our sample appear to be dusty, with $A_{V} \sim 0.1$--$1.1$\,mag.  However, the best-fit clumpiness parameters ($q$) show that the ISM may in fact be enhancing the observed Ly$\alpha$~EWs, as 10/14 objects had best-fit $q$-values of $< 1$, implying the EWs we observe have been enhanced over the EWs intrinsic to the star-forming regions.  We thus concluded that ISM enhancement of the Ly$\alpha$~EW is a plausible explanation for the observed plethora of high-EW sources.  Additionally, the ubiquity of dust implies that LAEs are {\it not} the primitive first galaxies as previously thought.  Finally, while the LAE population appears to be predominantly young, there is a non-negligible fraction of older, more evolved sources.  The relation between these rarer, old LAEs and the younger, more common LAEs is an important topic for future research.

\section{What the Near Future Holds}
The previous section has illustrated that at the very least, LAEs are a complicated category of objects.  The cause of the large observed EWs among narrowband-selected LAEs is still not clear, though we have shown that EW enhancement via a clumpy ISM is consistent with some observations, and thus is a plausible scenario.  However, it is clear that {\it on average}, LAEs are young ($< 100$\,Myr) and low mass ($< 10^{9} M_{\odot}$).  Additionally, studies of LAEs from $3 < z < 6$ \citep[e.g.,][]{gawiser06b, pirzkal07, finkelstein09a} show that their physical properties do not appear to change with redshift, including their physical size (Malhotra et al.\ 2010, in prep).  This is contrary to LBGs, which get larger \citep{ferguson04} and more massive \citep[e.g.,][]{stark09, finkelstein09f} with decreasing redshift.  This suggests a scenario where, at every redshift, LAEs are the building blocks of more evolved galaxies at lower redshifts \citep[e.g.,][]{gawiser07}.

In order to test this theory and truly understand the physical make-up of LAEs, we need to directly measure their properties.  For example, dust extinction is one of the least constrained properties deduced from SED fitting.  Fortunately, the UV energy that the dust absorbs is emitted in the rest-frame far-infrared (FIR).  Unfortunately, at high-redshift this emission is shifted into the sub-millimeter--millimeter regime.  In the future, the Atacama Large Millimeter Array (ALMA) should be able to detect the dust emission from $\sim 50$\% of LAEs in a relatively short amount of time \citep*{finkelstein09b, dayal09}, but this is years away.  

Currently, near-infrared (NIR) spectroscopy can resolve some of these issues.  For example, the Balmer decrement provides a direct measurement of the dust extinction.  For this experiment, one measures H$\alpha$ and H$\beta$ emission.  In the absence of dust, the ratio of H$\alpha$/H$\beta$ should be 2.86 \citep{osterbrock89}.  However, dust will absorb H$\beta$ more efficiently than H$\alpha$, thus any dust present will increase this value.  Comparing the measured value of H$\alpha$/H$\beta$ to the theoretical value allows one to measure the amount of dust attenuation.  However, H$\alpha$ measurements are currently restricted to $z \leq 2.7$ for practical reasons, as at higher redshifts this line shifts out of the K-band and is out of reach from the ground.  Currently, H$\alpha$ measurements of high-redshift LAEs exist for only a handful of objects from the Double-Blind survey (Hayes et al.\ 2009).  Thus, to truly probe the physical properties of LAEs, we need a large sample at $z < 2.7$.
\begin{figure}[~ht]
\centering
\includegraphics[width=5in]{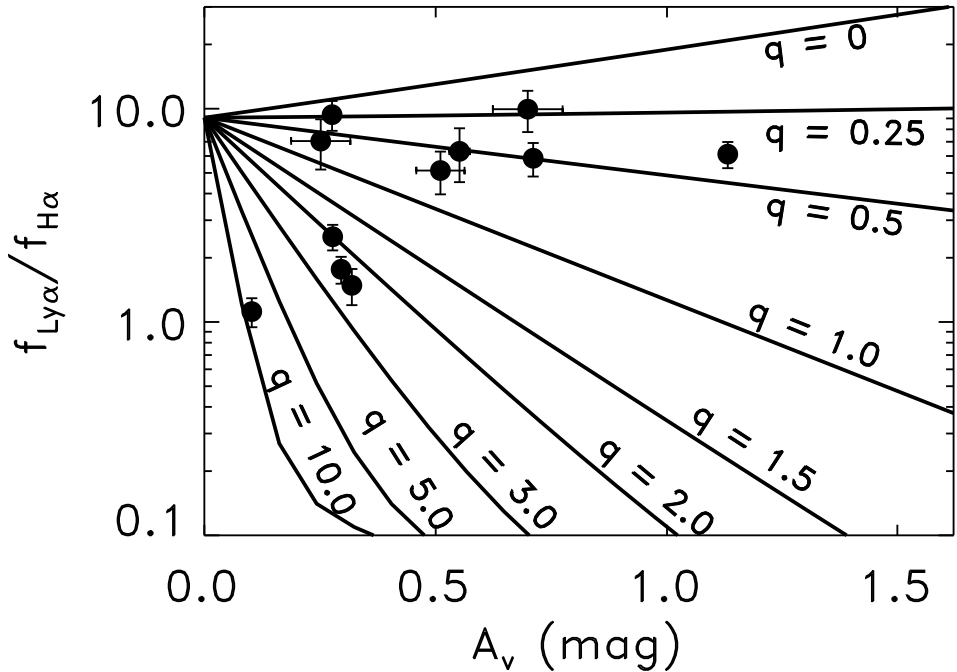}
\caption{The ratio of Ly$\alpha$/H$\alpha$ versus the measured dust extinction from the Balmer decrement for 11 LAEs at $z \sim 0.3$, confirmed to be star-forming \citep{finkelstein09c, finkelstein09e}.  The different curves show how Ly$\alpha$/H$\alpha$ changes with increasing dust extinction levels for various values of the ISM clumpiness parameter $q$.  Thus, it appears that at least for these low-redshift LAEs, the inhomogeneity of the ISM may be enhancing their Ly$\alpha$~EWs over the intrinsic value, as they are consistent with $q < 1$ (from Finkelstein et al. 2010, in prep).  NIR spectroscopy will allow similar investigations into LAEs at $z \sim 2$--3.}\label{qfig}
\end{figure}

Beginning in the fall of 2011, the Hobby Eberly Telescope Dark Energy Experiment \citep[HETDEX,][]{hill08} will provide these objects.  HETDEX will discover nearly one million LAEs at $1.9 < z < 3.5$ over $\sim 500$\,deg$^{2}$, in order to measure baryonic acoustic oscillations, and thus probe the evolution of the dark energy equation of state.  Measuring the rest-frame optical emission lines from the LAEs at the lower end of this redshift range (possible in a few hours with 8--10m class telescopes) will provide some of the first direct measurements of the dust extinction and the metallicities of LAEs.  Additionally, via ratios of rest-frame optical emission lines \citep*{baldwin81}, one can also probe for AGN contamination, as low-luminosity AGN may be lurking in the shadows.  Finally, by comparing the ratio of Ly$\alpha$~to a non-resonant line (i.e.~H$\alpha$), one can directly probe the geometry of the ISM \citep[cf.][]{atek09, scarlata09}.  Figure~\ref{qfig} shows 11 GALEX-discovered LAEs at $z \sim 0.3$ which were determined to be star-formation dominated \citep{finkelstein09e}.  In a uniform ISM, the ratio of Ly$\alpha$/H$\alpha$ should decrease rapidly with increasing dust extinction, as shown by the $q=10$ curve.  However, as the ISM becomes more clumpy, Ly$\alpha$~becomes less attenuated with respect to H$\alpha$ (and the continuum), as shown by the curves for decreasing values of $q$, until at very small values of $q$, the ratio of Ly$\alpha$/H$\alpha$ becomes greater than the intrinsic, theoretical value of~$\sim 9$ \citep{osterbrock89}.  Using optical spectroscopy of these low-redshift LAEs, we determined that many of them appear to have ISMs which enhance their observed Ly$\alpha$~EWs (Finkelstein et al.\ 2010, in prep), consistent with their stellar population modeling results \citep{finkelstein09c}.  At high redshift we do not currently have any direct probes into the state of the ISM, but rest-frame optical emission lines will provide these probes in the near future.

\section{Approaching the Big Bang}
The future of galaxy evolution studies diverges into two extremes.  As described above, we need to study LAEs (and all galaxy types) at $z < 3$ in order to directly measure their physical properties via diagnostic emission lines.  At the same time, pushing galaxy studies to the highest redshifts possible can tell us more about how the first galaxies formed, and their impact on the last major phase transition of the Universe, the reionization of the IGM.  Perhaps the best chance of discovering galaxies at $z \geq 7$ rests with the narrowband selection technique, as the bright Ly$\alpha$~emission can compensate for the very faint continuum levels.  At these redshifts, Ly$\alpha$~is observed in the NIR, thus strong atmospheric OH emission causes difficulties.  The typical way to compensate for this is to manufacture ultra-narrowband filters (FWHM $\sim 10$--100\,\AA) which are sensitive to regions between these sky lines.  The most promising gaps are at $\sim 1.06$ and 1.19 $\mu$m, corresponding to Ly$\alpha$~at z = 7.7 and 8.8.  Many ongoing or planned surveys will exploit these gaps, such as DAzLE \citep{horton04} and ELVIS \citep{nilsson07b}.  Currently, \citet{hibon09} found 7 $z \sim 7.7$ LAE candidates down to f$_{{\rm Ly}\alpha} \sim 8\times 10^{-18}$\,erg s$^{-1}$ cm$^{-2}$ with a 40 hour 1.06 $\mu$m narrowband exposure with the CFHT.  Using a 28 hour exposure obtained with the NEWFIRM wide-field NIR camera on the KPNO 4m, Tilvi et al.\ (2010, in prep) discovered another few candidate $z \sim 7.7$ LAEs.  However, all but one (from \citealt{hibon09}) of these candidates were only detected in the narrowband and not in any broadbands, although this is not unexpected given the limitations of ground-based NIR imaging.  Thus, we require spectroscopic follow-up to determine the validity of these candidates.
\begin{figure}[!ht]
\centering
\includegraphics[width=5in]{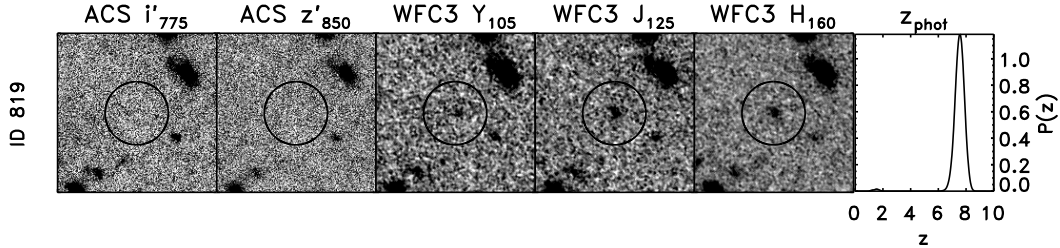}
\caption{5{\hbox{$^{\prime\prime}$}}~cutouts of a $z_{phot} = 7.55$ galaxy discovered with WFC3, from \citet{finkelstein09f}.  The right-hand plot shows the probability distribution function (PDF) of the photometric redshift.  This object has $> 99$\% of its PDF at $z > 6$.}\label{lbgstamp}
\end{figure}

At $z \sim 3$--4, it is straightforward to obtain large samples of LBGs and LAEs.  At $z \sim 5$--6, the dropping observed continuum brightness makes the LAE selection more powerful.  However, at $z \geq 7$, we've gotten to the point where ground-based narrowband selection is becoming extremely difficult.  Due to the recently installed Wide Field Camera 3 (WFC3) on board \textit{HST}, we now have the ability to probe down to $m_{\rm AB} \sim 29$ in the NIR with broadband photometry, thus LBG selection at the highest redshifts is now \textit{easier} than narrowband LAE selection.  Soon after WFC3 was commissioned, it obtained 60 orbits in the HUDF in the $Y$, $J$ and $H$-bands, at 1.05, 1.25 and 1.6 $\mu$m, respectively (PI Illingworth).  \citet{oesch09} and \citet{bouwens09} published the discoveries of 15 $z \sim 7$ (z$^{\prime}$ dropout) and 5 $z \sim 8$ ($Y$ dropout) LBG candidates.  Among their conclusions were that these galaxies were very blue \citep{bouwens09c}.  Does this imply that we are finally observing the first galaxies?
\begin{figure}[!ht]
\centering
\includegraphics[width=5in]{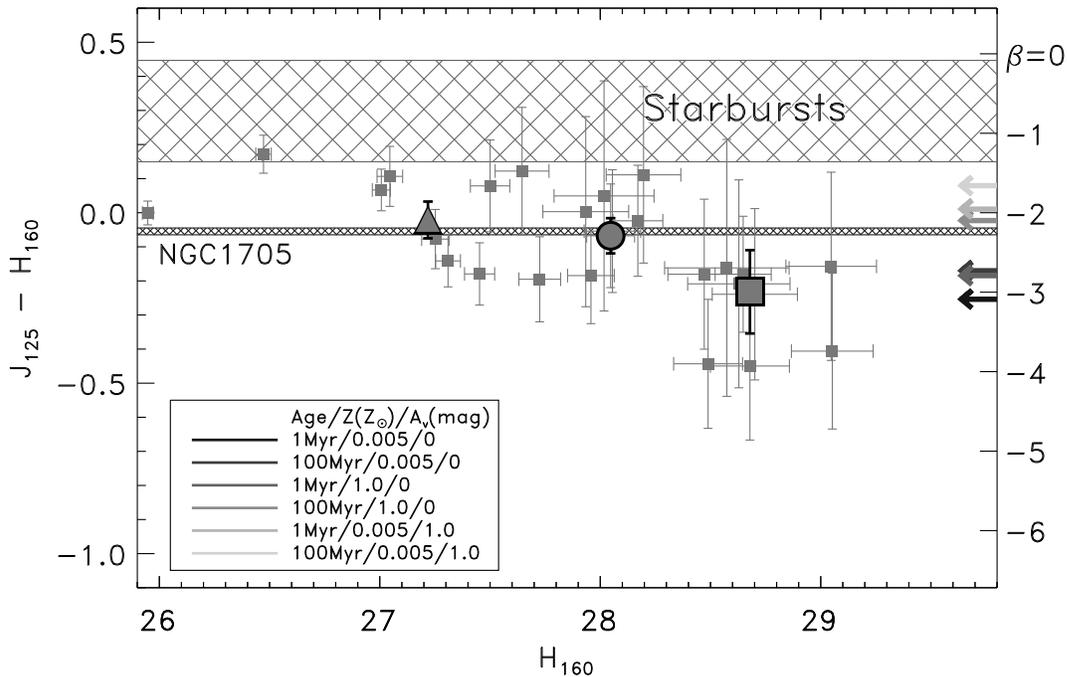}
\caption{A color-magnitude diagram for galaxies at $6.3 < z < 7.5$, from \citet{finkelstein09f}.  The individual points are the small gray squares.  We computed the average colors for all objects (circle), objects with $H < 28.5$ (triangle) and $H \geq 28.5$ (large square).  While LBGs at $z \sim 3$ have colors consistent with local starbursts, all of these bins are significantly bluer.  The faint bin is bluer than even NGC1705, which is one of the bluest local starbursts known.  This implies that these galaxies, especially the faint ones, are consistent with having young, low metallicity and relatively unextincted stellar populations.}\label{colmag}
\end{figure}

To take this one step further, we used an updated reduction of the WFC3 data to perform an independent photometric redshift analysis \citep[similar to][]{mclure09}, finding 35 candidate galaxies at $6.3< z < 8.6$ \citep{finkelstein09f}.  Figure~\ref{lbgstamp} shows one of our galaxy candidates at $z_{\rm phot} = 7.55^{+0.34}_{-0.36}$.  We examined the rest-frame UV colors of these galaxies, comparing them to the empirical local starburst templates of \citet{kinney96} as well as a number of synthetic stellar populations using updated models from Bruzual \& Charlot (2003).  Figure~\ref{colmag} shows the $J - H$ colors of the subsample of galaxies at $z < 7.5$ versus their $H$-band magnitudes.  Given that the photometric errors on the individual objects are rather large, we computed an average color for all objects (circle), objects with $H < 28.5$ (triangle) and objects with $H \geq 28.5$ (large square).  These means and their corresponding errors were computed via bootstrap simulations, where in each simulation we varied the color of the individual points by an amount proportional to their photometric scatter (which we determined with separate simulations), and then recomputed the mean color.

It is immediately apparent that both the bright and the faint galaxies are significantly ($>4\sigma$) bluer than local starbursts.  This is interesting, as \citet{papovich01} performed a similar analysis on LBGs at $z \sim 3$, and found their colors to be {\it consistent} with the Kinney et al.~templates.  Rather, these galaxies are consistent with NGC1705, which is one of the bluest local starburst known.  The stellar population models, shown as the shaded arrows on the right-hand side, imply that these bluest galaxies are consistent with young ages ($< 100$\,Myr) sub-solar metallicities and little-to-no dust extinction, representing a change from the more evolved LBGs typical at $z < 6$.
\begin{figure}[!ht]
\centering
\includegraphics[width=2.5in]{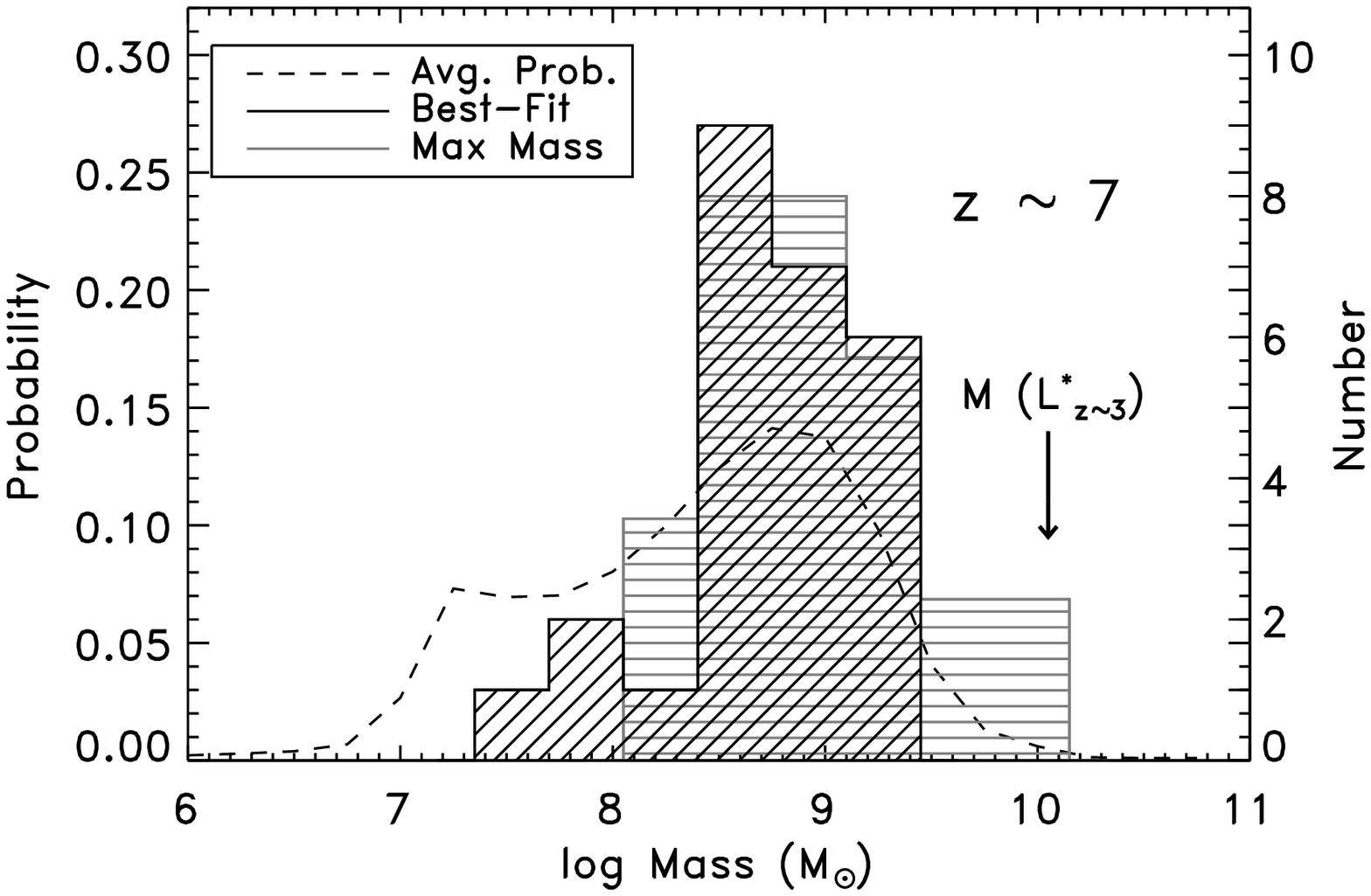}\label{z7mass}
\includegraphics[width=2.5in]{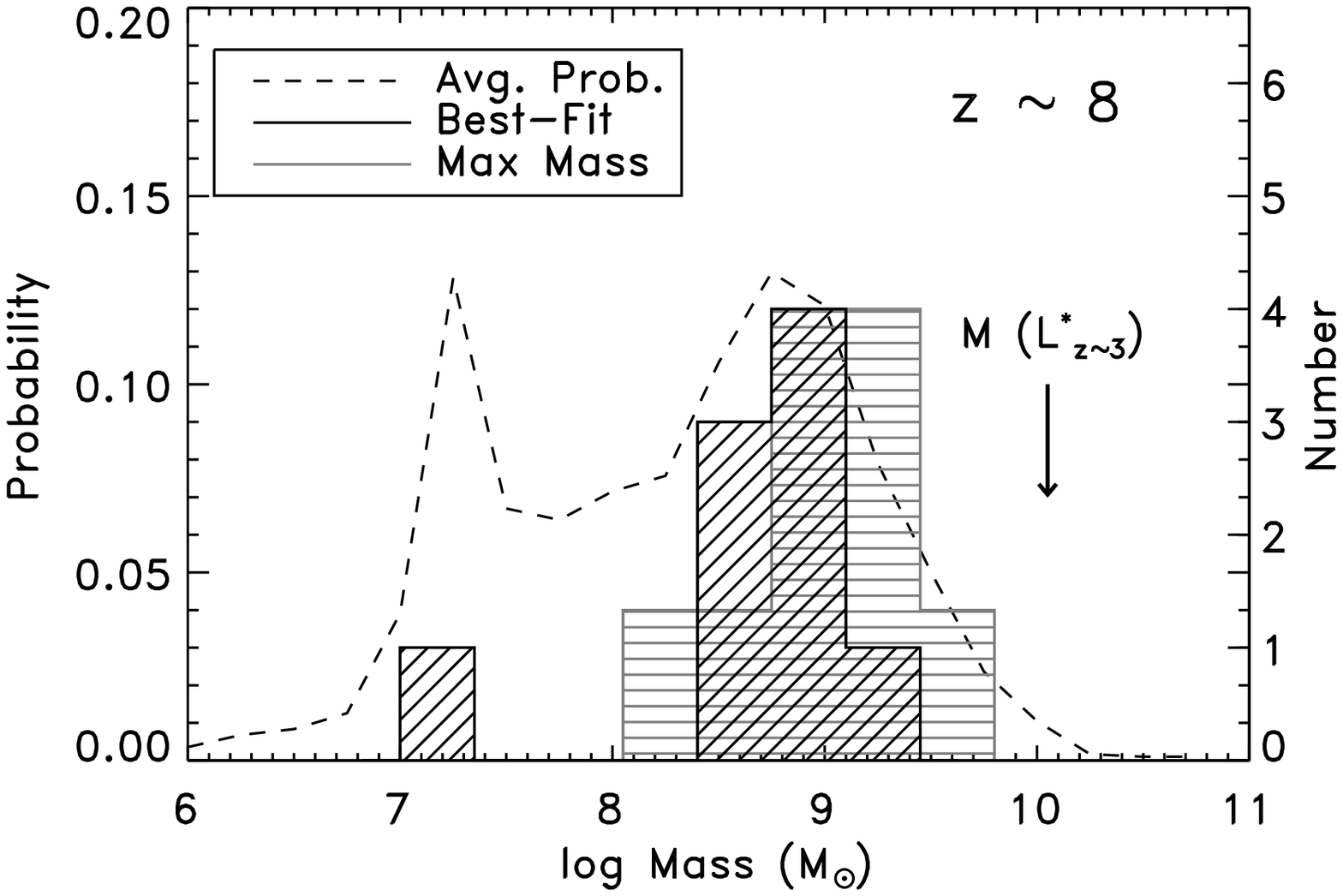}\label{z8mass}
\caption{The best-fit and maximum masses for our sample at $z \sim 7$ (left) and $z \sim 8$ (right), showing a typical stellar mass of 10$^{8}$--10$^{9}$ $M_{\odot}$.  These galaxies are lower mass than $z \sim 3$ LBGs (shown by the arrow), and we even see a hint of evolution from $z \sim 7$, as $z \sim 8$ LBGs are more likely to have $M \sim 10^{7} M_{\odot}$.}\label{massdist}
\end{figure}

To learn in more detail about the physical properties of these highest redshift objects, we performed SED fitting using their observed WFC3 fluxes, as well as limits from ACS and IRAC.  With only a few detected data points, definitive determinations of their ages, dust extinction and metallicities are difficult.  However, they do appear to be young, consistent with the color analysis above, as 30/35 objects have very young ages of $< 10$\,Myr within the 68\% confidence range on their age (as determined from 10$^{3}$ bootstrap simulations).  While the mean extinction is $A_{V} \sim 0.4$\,mag, over half of the sample is consistent with no dust extinction.  Lastly, while the metallicity is traditionally difficult to decipher in SED-fitting analysis, the very blue colors of these objects allow us to place a significant limit on the metallicity, with $Z < 0.1 Z_{\odot}$~significant at the $1\sigma$ level.
\begin{figure}[!ht]
\centering
\includegraphics[width=5in]{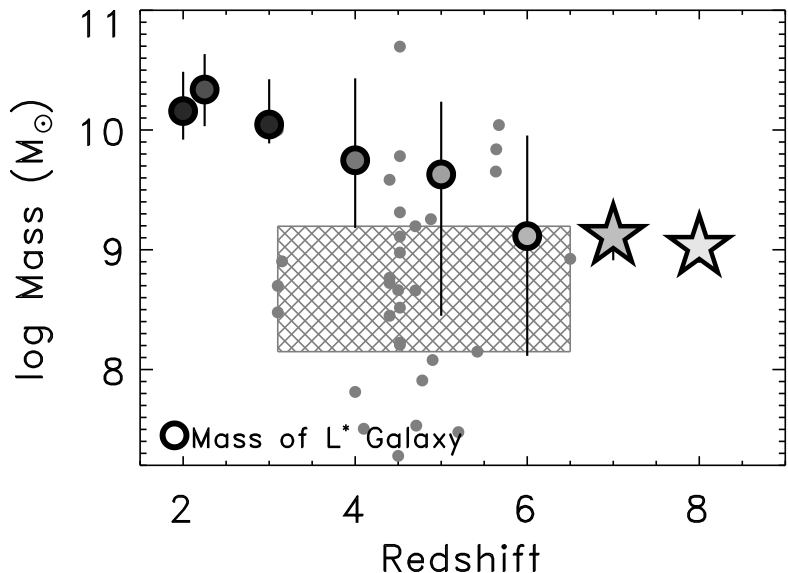}
\caption{The stellar masses of $L^{\ast}$ galaxies versus redshift from Finkelstein et al.\ (2009d), with comparison samples taken from the literature.    Our results are the large stars at $z \sim 7$ and $z \sim 8$.  The remaining symbols, from dark to light,  are: \citet{reddy06}, \citet{shapley05} and \citet{stark09} at $z = 4$, 5 and 6, respectively.  The circles denote the mass at $L^{\ast}$ (characteristic luminosity) at each redshift, using the luminosity functions from \citet{bouwens07}, \citet{reddy09} and \citet{oesch09}.   The background gray circles denote stellar masses of Ly$\alpha$~emitting galaxies at $3.1 \leq z \leq 6.5$, taken from: \citet*{chary05}, \citet{gawiser06b}, \citet{pirzkal07}, \citet{nilsson07a}, \citet{lai07}, \citet{finkelstein07}, \citet{lai08} \& \citet{finkelstein09a}.  The gray hatched region denotes the interquartile range of the LAE masses.  The masses of the  $z \geq 6.3$ LBGs studied here are more similar to those of LAEs at all redshifts than LBGs at any redshift $< 6$.}\label{masses}
\end{figure}

While the above properties are difficult to directly pin down as they all result in reddening of the SED, constraints on the stellar masses are more robust, as the degenerate combination of all other parameters yields similar mass-to-light ratios.  Figure~\ref{massdist} displays the distribution of best-fit masses at $z \sim 7$ and $z \sim 8$, as well as the cumulative probability distribution of stellar masses at each redshift.  This figure shows that the galaxies in our sample have typical masses of 10$^{8}$--10$^{9}$ $M_{\odot}$, with galaxies at $z \sim 8$ having a higher probability of having masses of $<10^{8} M_{\odot}$.  However, these masses were derived using solely rest-frame UV fluxes, which can miss significant amounts of mass in older stars lost beneath the glare of the UV-bright young stars.  We thus performed a second fit, allowing two bursts of star formation, where one burst was constrained to occur at $z = 20$, and contain 90\% of the stellar mass.  We refer to the best-fit masses from these fits as the maximum mass, as they contain the maximum physically plausible amount of mass which could exist in these galaxies without violating the {\it Spitzer} flux limits (although physically unlikely, we did perform a fit with 99\% of the mass in old stars, and found the maximum mass to only increase by a factor of $\sim 2$).  Investigating Figure~\ref{massdist}, one can see that these maximum masses are only shifted to marginally higher masses than the best-fit masses, implying that the {\it Spitzer} upper limits are providing little constraints on the stellar masses.  This is understandable, as at these redshifts, the time elapsed since $z = 20$ is only $\sim 400$--700\,Myr, thus late-type B/early-type A stars will still be on the main sequence, and these stars contribute measurably to the rest-frame UV.

Figure~\ref{masses} shows the masses of $L^{\ast}$ galaxies as a function of redshift, with the lower redshift points coming from \citet{reddy06}, \citet{shapley05} and \citet{stark09}.  Investigating this figure, from $z = 2$--6 the mass of a typical $L^{*}$ galaxy appears to drop by an order of magnitude, though the spread in the $z \sim 6$ point made an exact determination difficult.  With the addition of our data points, we now confirm this trend of decreasing mass with increasing redshift, with a slight (though not significant) drop in mass from $z \sim 7$--$8$.  The small gray circles denote the masses of LAEs taken from the literature, and the hatched box denotes their interquartile range.  At $z \leq 5$, LBGs are systematically more massive than LAEs.  However, at $z \geq 6$, there is a change in the physical properties of typical continuum selected galaxies, and their masses are more consistent with LAEs at all redshifts, than with LBGs at any previous redshift.  Given the blue colors and corresponding un-evolved characteristics of the $z \sim 7$--8 galaxies discussed here, it appears that at these high redshifts, evolved LBGs common at lower redshifts no longer exist, and our sample consists of galaxies which are the true building blocks of lower-redshift evolved galaxies, similar to LAEs at lower redshifts \citep[e.g.,][]{gawiser07}.

\section{Conclusions}
Selecting galaxies on the basis of their Ly$\alpha$~emission allows the discovery of a plethora of faint, high-redshift galaxies.  Numerous studies over the past few years have shown that these galaxies primarily appear to be a unique population of star-forming galaxies, which are young ($< 100$\,Myr) and low mass ($< 10^{9} M_{\odot}$), although a minority can be older and more massive.  Many of the LAEs which have been discovered exhibit strong Ly$\alpha$~equivalent widths which cannot be explained by normal stellar populations.  While AGN contamination could reconcile this problem, AGN (at least luminous ones) appear to be rare among LAEs.  Very low metallicities could also cause high EWs, but the dustiness of many LAEs likely rules out this scenario.  Finally, this dust, if in a clumpy geometry, can enhance the Ly$\alpha$~EW by increasing the escape fraction of Ly$\alpha$~relative to the continuum, and this scenario can explain the SEDs of a sample of high-redshift LAEs.  However, to truly determine the make-up of LAEs, including their dust properties, we require direct measurements, which is available presently with NIR spectroscopy, and in the near future with ALMA.

Pushing galaxy studies to the highest redshifts, ground-based narrowband selection of LAEs becomes extremely difficult due to the bright NIR sky background, with the deepest studies finding only a handful of candidate $z > 7$ LAEs.  However, the addition of WFC3 to {\it HST} provides the capability to obtain large samples of $z > 7$ galaxy candidates.  Initial studies of these galaxies have shown them to be somewhat primitive in their make-up, with young ages ($< 100$\,Myr), low masses ($10^{8}$--$10^{9}$ $M_{\odot}$), low metallicity ($< 0.1Z_{\odot}$), and consistent with little-to-no dust extinction.  Comparing the physical properties of these most distant galaxies to those at lower redshifts, they appear more similar to LAEs at all redshifts than LBGs at any redshift.  This implies that typical galaxies at $z > 7$ are less evolved than at lower redshifts.  At some point in the past we should begin to see the first building blocks of galaxies, and these observations show that $z=$7--8 may be this epoch, although larger samples of spectroscopically confirmed galaxies will be necessary to make a definitive determination.

\acknowledgements S.\ L.\ F.\ would like to thank my collaborators on these projects: James Rhoads, Sangeeta Malhotra, Casey Papovich, Norman Grogin, Nor Pirzkal, Junxian Wang, Harry Ferguson, Mark Dickinson, Mauro Giavalisco, Naveen Reddy and Anton Koekemoer.  S.\ L.\ F.\  would also like to thank Keely Finkelsten for her help with preparing the talk, and Keely Finkelstein and James Rhoads for a careful read through of these proceedings.  S.\ L.\ F.\  is supported by the Texas A\&M University Department of Physics and Astronomy.

\end{document}